\documentstyle[epsfig,11pt]{article}
\newcommand{\spin}{{\sc Spin}}
\newcommand{\promela}{{\sc Promela}}
\newcommand{\otter}{{\sc Otter}}
\newcommand{\imp}{{\longrightarrow}}
\begin{document}
\pagenumbering{roman}
\thispagestyle{empty}
\setcounter{page}{1}
\begin{center}

\vspace{.2in}
\rule{1.5in}{.01in}\\ [1ex]
ANL/MCS-TM-261 \\
\rule{1.5in}{.01in}

\vspace{1.5in}
{\Large\bf Methods to Model-Check Parallel Systems Software}\\[2ex]
by \\ [2ex]

{\Large\it Olga Shumsky Matlin, William McCune, and Ewing Lusk}\\
        {\tt{\{matlin,mccune,lusk\}@mcs.anl.gov}}

\vspace{1.5in}
Mathematics and Computer Science Division

\bigskip

Technical Memorandum No. 261

\vspace{1in}
April 2003
\end{center}

\vfill
\noindent
This work was supported by the Mathematical, Information, and Computational 
Sciences Division subprogram of the Office of Advanced Scientific Computing 
Research, Office of Science, U.S. Department of Energy, under Contract 
W-31-109-ENG-38.
\newpage
\noindent
Argonne National Laboratory, with facilities in the states of Illinois
and Idaho, is owned by the United States Government and operated by The
University of Chicago under the provisions of a contract with the
Department of Energy.

\vspace{2in}

\begin{center}
{\bf DISCLAIMER}
\end{center}

\noindent
This report was prepared as an account of work sponsored by an agency
of the United States Government.  Neither the United States Government
nor any agency thereof, nor The University of Chicago, nor any of
their employees or officers, makes any warranty, express or implied,
or assumes any legal liability or responsibility for the accuracy,
completeness, or usefulness of any information, apparatus, product, or
process disclosed, or represents that its use would not infringe
privately-owned rights.  Reference herein to any specific commercial
product, process, or service by trade name, trademark, manufacturer,
or otherwise, does not necessarily constitute or imply its
endorsement, recommendation, or favoring by the United States
Government or any agency thereof.  The views and opinions of document
authors expressed herein do not necessarily state or reflect those of
the United States Government or any agency thereof.
\vspace{1in}
\begin{center}
Available electronically at http://www.doe.gov/bridge

Available for a processing fee to U.S. Dept.\ of
Energy and its contractors, in paper, from:

U.S. Department of Energy\\
Office of Scientific and Technical Information\\
P.O. Box 62\\
Oak Ridge, TN 37831-0062\\
phone: (865) 576-8401\\
fax: (865) 576-5728\\
email: reports@adonis.osti.gov
\end{center}
\newpage

\tableofcontents

\newpage

\setcounter{page}{0}
\pagenumbering{arabic}
\begin{center}
{\Large\bf Methods to Model-Check Parallel Systems Software}\\ [1ex]
by \\ [1ex]
Olga Shumsky Matlin, William McCune, and Ewing Lusk
\end{center}

\addcontentsline{toc}{section}{Abstract}
\abstract{We report on an effort to develop methodologies for formal
  verification of parts of the Multi-Purpose Daemon (MPD) parallel
  process management system.  MPD is a distributed collection of
  communicating processes.  While the individual components of the
  collection execute simple algorithms, their interaction leads to
  unexpected errors that are difficult to uncover by conventional
  means.  Two verification approaches are discussed here: the standard
  model checking approach using the software model checker \spin\ and
  the nonstandard use of a general-purpose first-order
  resolution-style theorem prover \otter\ to conduct the traditional
  state space exploration.  We compare modeling methodology and
  analyze performance and scalability of the two methods with respect
  to verification of MPD.}   

\section{Introduction}
\label{sec:introduction}
Reasoning about parallel programs is surprisingly difficult.  Even
small parallel programs are difficult to write correctly, and an
incorrect parallel program is equally difficult to debug, as we
experienced while writing the Multi-Purpose Daemon (MPD), a process
manager for parallel programs
\cite{bgl00:mpd:pvmmpi00,butler-lusk-gropp:mpd-parcomp}.  Despite
MPD's small size and apparent simplicity, errors have impeded progress
toward code in which we have complete confidence.  Such a situation
motivates us to explore program verification techniques.  

MPD is itself a parallel program.  Its function is to start the
processes of a parallel job in a scalable way, manage input and
output, handle faults, provide services to the application, and
terminate jobs cleanly.  MPD is the sort of process manager needed to
run applications that use the standard
MPI~\cite{mpi-forum:journal,mpi-forum:mpi2-journal} library for
parallelism, although it is not MPI specific.  MPD is distributed as
part of the portable and publicly available
MPICH~\cite{gropp-lusk:mpich-www,gropp-lusk-doss-skjellum:mpich}
implementation of MPI. 

Our first attempt to use formal verification to ensure correctness of
MPD algorithms \cite{mpd-acl2} employed the ACL2 \cite{acl2:book}
theorem prover, which allows one to both simulate and verify a model
within a single environment. Components of the MPD system,
as well as the elements of the Unix socket library on which MPD is
based, were formalized in a subset of Lisp.  The formalization was based on
a so-called oracle \cite{acl2:oracle}, which allows analysis of a
parallel system in a sequential environment.  The oracle specifies an 
execution interleaving of concurrent processes and is randomly
generated for simulations.  Verification is conducted with respect to
an arbitrary oracle (i.e., an arbitrary execution interleaving); 
thus, a property proved in such a way holds for all possible
executions of a collection of concurrent processes.  In this approach
parsing simulation results, formulating desired properties of models
of MPD algorithms and reasoning about such models proved difficult,
leading us to abandon this traditional theorem-proving method of
verification and to try instead model-checking techniques.  Two such
techniques are described here.

The first technique employed the model checker
\spin~\cite{spin:book,spin:article}, which
supports design and verification of asynchronous distributed
communicating systems.  Models of such systems are recorded in the
special high-level verification language \promela, which can also be
used to state some correctness properties of the models.  Additional
correctness properties are specified by using linear temporal logic.
The verification engine of \spin\ is based on on-the-fly reachability
analysis with several optimizations and heuristics, such as
partial-order reduction and bitstate hashing.  The system also
includes a simulation environment and a graphical user interface.
\spin\ has been used in various settings (see~\cite{spin-web-page} for
all of the proceedings of the \spin\ workshops).  Because the dynamic
nature of MPD is easily expressible in the \spin/\promela\ framework,
the tool is a natural choice for verification of our system.
However, our early experiences with \spin\ \cite{spin-mpd} suggest
that the most natural formalization of certain MPD algorithms in
\promela\ leads to performance and scalability challenges.  While we
have since addressed some of these challenges, as described in
Section~\ref{sec:spin}, they led us to explore other ways to formalize
and model check MPD algorithms.  

Specifically, because of considerable in-house theorem-proving 
expertise, we were led to explore whether a theorem prover can be
successfully adapted for our purposes.  We have used the
general-purpose first-order resolution and paramodulation theorem
prover \otter\ \cite{McCune94a}. While the tool is widely used, its
primary application is in proof search, mainly in mathematics and
logic.  The input language is that of first-order logic.  Otter does,
however, contain extensions for evaluable terms, which are essential
to our unusual use of the theorem prover for state-space exploration.

The remainder of the paper is structured as follows.  In
Section~\ref{sec:mpd} we describe MPD in more detail and briefly
present the algorithms, along with their correctness properties, for
which the verification approaches are evaluated.  In
Sections~\ref{sec:spin} and \ref{sec:otter} we outline formalizations
of the MPD algorithms in \promela\ and the input language of \otter. In
Section~\ref{sec:results} we compare and analyze concrete results of
specific verification experiments.  We conclude with a summary in
Section~\ref{sec:summary}.  

\section{The Multi-Purpose Daemon}
\begin{figure*}%
    \centerline{ \epsfxsize=4.75in \epsfbox{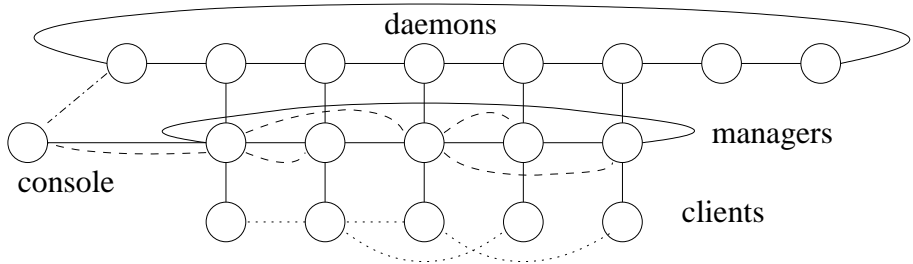}}
    \caption{Daemons with console process, managers, and clients}
    \label{fig:mpds-all}
\end{figure*}

The MPD system comprises several types of processes.  The {\em
 daemons\/} are persistent (may run for weeks or months at a time,
starting many jobs) and are connected in a ring.  {\em Manager\/}
processes, started by the daemons to control the application
processes ({\em clients\/}) of a single parallel job, provide most of
the MPD features and are also connected in a ring.  A separate set of
managers supports an individual process environment for each user
process.  A {\em console\/} process is an interface between a user
and the daemon ring.  A representative topology of the MPD system is
shown in Figure~\ref{fig:mpds-all}.
\label{sec:mpd}
The vertical solid lines represent
connections based on pipes; the remaining solid lines represent
connections based on Unix sockets.  The dashed and dotted lines are
potential or special-purpose connections.

Each of the daemon, manager, and console process types has essentially
the same pattern of behavior.  This feature important because it
allows us to model these processes in a consistent manner.  After 
initialization, the process enters an infinite, mostly 
idle, loop, implemented by the Unix socket function \texttt{select}.
When a message arrives on one of its sockets, the process calls the
appropriate message handler routine and reenters the idle {\tt
 select} state.   The handler does a small amount of processing,
creating new sockets or sending messages on existing ones.  The 
logic of the distributed algorithms executed by the system as a whole
is contained primarily in the handlers, and this is where the
difficult bugs appear.     

In the following we compare the two verification approaches on two
algorithms: a daemon-level ring establishment algorithm and a
manager-level barrier algorithm.  We present an outline of each  
algorithm and highlights of the modeling approach. 

\subsection{Ring Establishment Algorithm}
\begin{figure*}%
    \centerline{ \epsfxsize=4.75in \epsfbox{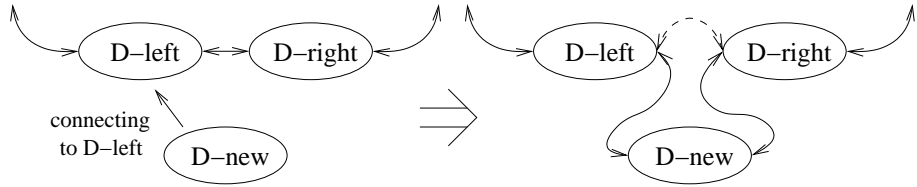} }
    \caption{Ring insertion}
    \label{fig:insert}
\end{figure*}

Establishment and maintenance of a ring of daemons are central to the
operation of MPD.   Informally, daemon ring creation proceeds as
follows.  The initial daemon establishes a listening port to which
subsequent connections are made.  The daemon connects to its own
listening port, creating a ring of one daemon.  The listening port of
the first daemon and the name of the host processor are queried from
the console.  The desired number of daemons is then initiated and
directed  to enter the ring by connecting to the first daemon.
Figure~\ref{fig:insert}   
shows the result of inserting a new daemon into an existing ring.
After the insertion is completed, the old connection between
daemons on the right and left of the new daemon is disconnected (shown
in the figure by the dashed line).  Note that in the special case of
insertion into a ring of one daemon, the daemon plays both the left
and the right roles.  

The algorithm may be viewed as consisting of two parts, each of which
may potentially contain errors.  One part deals with establishment of
listening ports and sockets to enable bidirectional communication
between processes.  The second part concerns the passage of messages
over the established communication links and their handling upon
receipt.  Correct socket establishment depends to a large degree on
the correct implementation of the Unix socket library.   The \promela\
and \otter\ models described below do not check correctness of this
part of the algorithm but rather concentrate on the correctness of
passing messages between processors, correctness of message handling,
and correctness and consistency of the global system state.  Under
this partitioning of the algorithm, and assuming that each processor
records the identity of its right and left neighbors, the correctness
property, stated as a postcondition, can be formulated as follows.

The algorithm is correct for $i$ daemons when, upon termination of the
algorithm, the ring has $i$ distinct connected components and when, for
any processors $i$ and $j$ in the ring, if the right neighbor of $i$
is $j$, then the left neighbor of $j$ is $i$. 

\subsection{Barrier Algorithm}

Parallel programs frequently rely on a \textit{barrier} mechanism to
ensure that all processes of the job reach a certain point (complete
initialization, for example) before any are allowed to proceed
further.  Parallel jobs, that is, programs running on the clients, rely
on the manager processes to implement the barrier service.  The
algorithm proceeds as follows.  A manager process is designated as
the leader of the algorithm and is given a rank of 0.   When the
leader reads a request from its client to provide the barrier service,
it sends a message \texttt{barrier\_in} to its right-hand side neighbor
in the ring.  When a non-leader manager receives the
\texttt{barrier\_in} message, its behavior is determined by whether its
client has already requested the barrier service.  If the client has
done so, the manager forwards the message to the right-hand side
manager.  Otherwise, it holds the \texttt{barrier\_in} message until
the request from the client arrives.  While the \texttt{barrier\_in}
message is held, a bit variable \texttt{holding\_barrier\_in} is set.
Once the \texttt{barrier\_in} message traverses the entire manager ring
and arrives back in the leader, meaning that each client has reached
the barrier and notified its manager, the leader sends a
\texttt{barrier\_out} message around the ring.  When a manager receives
the \texttt{barrier\_out} message, it notifies its client to proceed
past the barrier.  The leader can be either the first or the last
manager to allow its client to proceed.

The barrier algorithm, as the ring establishment algorithm, can be
viewed as consisting of two parts: socket handling and message
handling.  Here, however, the socket handling portion of the algorithm
is largely unimportant to the verification effort, since the
communication paths between processes are completely established
before the algorithm begins execution.  The socket portion may become
important in future verification efforts that concentrate on
\textit{interaction} of MPD algorithms and examine the barrier
algorithm in conjunction with an algorithm that manages --- that is,
establishes, breaks down, or restores after a fault --- connections 
between processes.

Two correctness properties, a postcondition and an invariant, are
verified for the barrier algorithm.  The postcondition property is that
all clients have been released from the barrier.  The invariant
demands that no client be released  until every client
has reached the barrier, that is, every client has requested the
barrier service from its manager.

\section{Formalization and Verification of MPD with \spin}
\label{sec:spin}

To formalize an MPD algorithm in \promela, the modeling language of
\spin, one has to make three related decisions: what variables
need be defined globally and locally to record the necessary
information during the execution of the algorithm, how to model
the communication network, and whether and how to model the Unix
socket functions.  When considered together, the three modeling
decisions determine how abstract the resulting model will be.    

The original approach \cite{spin-mpd} to model checking MPD algorithms
with \spin\ produced models that were too literal, meaning that the
\promela\ modeling language was used in a way that closely resembled
C.  The motivation for such an approach was to enable automated
extraction of executable C code from verified \promela\ models.  As a
consequence of such a modeling methodology it was possible to
formalize within a single model both the socket-handling and the
message-handling portions of the MPD algorithms.  Unfortunately,
another consequence of the approach was poor verification performance.
We were able to verify the ring establishment algorithm on only a few (
less than five) daemons, far short of the desired goal to verify
models with ten to twenty processors (see Section~\ref{sec:size-note}
for discussion of why verifying models of this size would be
interesting).  During verification attempts on 
larger models, the number of states in the search space was so large
that \spin\ ran out of nearly 1 GB of available memory.   The current
formalization approach produces more abstract models, which enables
verification to complete within the constraints of available memory on
larger models, but at the expense of not considering
correctness of the socket-handling portion of the algorithms.

Components of the MPD system map naturally to predefined {\sc Promela}
entities: a  
{\tt proctype} is defined for each different MPD process type;
individual daemon, manager, console, and client processes correspond to 
active instances of the corresponding {\tt proctype}s; sockets map to
channels; and messages that are read and written over the sockets
correspond to messages traveling on the channels.  The functionality of 
the Unix {\tt select} \cite{stevens-unp1} is implemented by the
nondeterministic {\tt do} construct: when a message arrives in the
input queue of a process instance, the appropriate guard of the {\tt
  do} construct triggers its handling; in the absence of new messages
in the input queue, the instance is blocked.

The distinguishing characteristic of the \promela\ model of the
barrier algorithm is that the clients are not represented by separate
process instances. To do otherwise would waste of precious memory. In
\spin, each process instance requires a certain amount of space in the
description of the entire system state, otherwise known as the state
vector.  The functionality of a client is to send a request message to
its manager and then wait for a reply.  Both these actions can be
adequately modeled by two binary variables, \verb2client_barrier_in2
and \verb2client_barrier_out2, per manager process.  We note that the
algorithm depends on the fact that a manager records whether it has
already received a barrier request from its client, so one of these
variables is present in the model regardless of how clients are
formalized.  Omission of clients reduces the complexity of the model
and the size of the state vector in another way: communication links
between managers and clients are not necessary.  Connections between
managers are the only communication links that have to be modeled.  To
this end we define a global vector of $N$ channels, where $N$ is the
number of managers in the model.  Each manager process has local
pointers, \texttt{left} and \texttt{right}, to the appropriate
channels in the global array.  To manage the bit variables, we rely on
an implementation of bit arrays by Ruys \cite{spin:low-fat}, which
defines the \texttt{IS\_0} and \texttt{SET\_1} macros.  The
formalization ensures that the handler of a barrier request message
from the client is executed only once.  The use of the {\tt do}
construct ensures that all messages that arrive at the manager have
equal precedence and can be handled in any order. We show a partial
\promela\ model of the barrier algorithm:
{\small 
\begin{verbatim}
end_select: 
do
:: IS_0(client_barrier_in,_pid) ->
   SET_1(client_barrier_in,_pid);
   /* remainder of client request handling */
:: left?barrier_in ->
   /* barrier_in handling */
:: left?barrier_out ->
   /* remainder barrier_out handling */
   SET_1(client_barrier_out,_pid);
   if
   :: _pid == 0 -> skip
   :: else -> right!barrier_out
   fi;
od       
\end{verbatim}}

The socket-based communication network presented the most challenges
in the formalization of the ring establishment algorithm.  Unlike the
static topology of the communication links in the barrier algorithm,
the topology of connections in the ring algorithm is dynamic.  As
daemons enter the ring, new connections are established, and the old
ones are broken down, as illustrated in Figure~\ref{fig:insert}. In
addition, because all execution interleavings are exhaustively
considered during the model search, the network has to be set up so
that any two processes can communicate with each other. Numerous
different formalizations of the communication network were tried
in our effort to verify a model of meaningful size.  In particular, we
tried the bus and matrix of channels formalizations, as suggested 
in \cite{phd:ruys}. 

\begin{figure}%
    \centerline{ \epsfxsize=1.6in \epsfbox{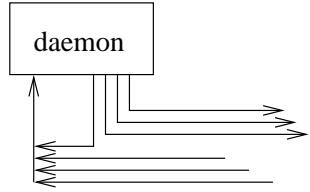} }
\caption{Model for message sending and receiving}
\label{fig:channels}
\end{figure}

Our experiments showed that, with respect to verification, the best
approach to modeling the communication network of the ring
establishment algorithm depends on a vector of global channels.  Each
daemon process owns a channel, which serves as an input queue for any
messages addressed to that process.  In order to further reduce
complexity, as suggested by \spin's documentation, each daemon retains
exclusive rights, by using the  \texttt{xr} designation, to read from
its owned channel.  In this approach, illustrated in
Figure~\ref{fig:channels}, 
a single input queue is used by a node to receive messages from all
nodes, but different queues are used to send messages addressed to
different nodes.  
We find it interesting that this approach is explicitly advised 
against in the \spin\ help file: to reduce complexity, the
documentation suggests one ``avoid sharing channels between multiple
receivers or multiple senders.'' This approach nonetheless succeeds in
our model because, while the network is set up to allow sharing
channels between multiple senders, most communications take place
between distinct pairs of daemons, so channel sharing very rarely occurs
during exploration of a single execution path. Thus, the complexity
of the model is not adversely affected.

\section{Formalization and Verification of MPD with \otter}
\label{sec:otter}

\otter\ \cite{ McCune94a} is an automated theorem-proving system for
first-order logic with equality.  Most successful applications of
\otter\ have been in abstract algebra and logic, and its strength is
the ability to quickly explore large search spaces.  It can
efficiently manage large sets of formulas (hundreds of thousands),
which allows it to automatically prove many difficult theorems.

\otter's basic operations are (1) to apply various inference rules
to formulas, (2) to apply rewrite rules to inferred formulas, and (3)
to determine whether newly inferred formulas are already in the
database of formulas.  These operations can be applied to
state space searches as well as to the heuristic searches
used in traditional automated theorem proving.  

Program verification is a nonstandard application of \otter.
Fortunately, \otter's language, data structures, and operations allow
reasonably intuitive and efficient implementations of model checking
by state space search.  One of the languages \otter\ accepts is a
sequent language with which we can write rules, assertions, and goals
as follows. 

\[
\begin{array}{l}
\mbox{\it hypothesis}_1, \cdots, \mbox{\it hypothesis}_n \imp conclusion.\\
\imp \mbox{\it assertion}.\\
\mbox{\it goal} \imp.
\end{array}
\]

\noindent
Given an input consisting of statements of this type, \otter\
can apply the rules to the assertions, generating new assertions,
and so on, until it derives one of the goals or until reaches a
fixed point (i.e., runs out of things to do).  All of the assertions
(initial and derived) are stored in the database.

A feature of \otter\ that allows it to conduct an efficient
state-space search is the ability to evaluate hypotheses by rewriting, 
as well as to match the hypotheses with assertions.  For example, the
hypotheses $X == 3$, where $X$ is instantiated by a preceding hypothesis,
is evaluated to a Boolean value.  The rewrite mechanism includes a
simple but general equational programming language so that the evaluable
hypotheses can be arbitrary function calls.  In addition,
the conclusion of the rule can have function calls to transform
the result.

One can use this mechanism to implement a state space search in
a straightforward way.  The initial states are initial assertions
(usually there is just one), say,
\[
\imp \mbox{\it State($s_0$)},
\]
where $s_0$ is a data structure representing the initial state.
The rules take states to successor states,
\[
\mbox{\it State(S)}, \mbox{\it P(S)} \imp \mbox{\it State(f(S))},
\]
where {\it P(S)} are evaluable hypotheses specifying the states to
which the rule applies, and {\it f(S)} generates the successor state.

When applying this method to model-checking concurrent computations,
one frequently needs the hypothesis ``Let $N$ be an arbitrary process.''
To implement this capability, we include initial assertions giving the 
set of process IDs, say, for a three-processor simulation,
\[
\begin{array}{l}
\imp  \mbox{\it PID(0)}.\\
\imp  \mbox{\it PID(1)}.\\
\imp  \mbox{\it PID(2)}.
\end{array}
\]
Then, the rules can be given as
\[
\mbox{\it State(S)}, \mbox{\it PID(N)}, \mbox{\it P(S,N)} \imp \mbox{\it State(f(S,N))}.
\]

In its basic form, the \otter\ search mechanism is designed to be
complete; that is, if a conjecture is a theorem, then \otter\ will
eventually find a proof.  (In the practice of traditional automated
theorem proving, restrictions are frequently imposed by the user,
resulting in incomplete search procedures.)  In the context of
state space search, this completeness means that (given enough
time and memory) all states that can be reached from the initial
states will eventually be derived and stored by \otter.

To determine whether an illegal state can be reached, one can include 
a rule such as
\[
\mbox{\it State(S)}, \mbox{\it P(S)} \imp \mbox{\it Bad\_State(S)}
\]
and the goal
\[
Bad\_State(S) \imp,
\]
with the effect that if any illegal state is derived, \otter\
will report it and print the path of states from an initial state to
the illegal one.  From \otter's point of view, it has simply found and
printed a proof of a goal.  From the point of view of verification,
the path of states represents an erroneous execution scenario.
The \otter\ proof of the reachability of an illegal state can be examined
to determine which state transition rule leads to the error. If \otter\
reaches a fixed point without deriving an illegal state, the terminal
states or the entire state space can be examined and processed by the
user or by another system. 

We show a 
partial three-element \otter\ model\ of the barrier algorithm:
{\small
\begin{verbatim}
%
-> PID(0).
-> PID(1).
-> PID(2).

%
-> State(PS(0,0,0,[]),PS(0,0,0,[]),PS(0,0,0,[])).

%
State(S), PID(X), X == 0, $TRUE(barrier_out_arrived(S,X)) ->
State(assign_client_return(receive_message(S,X),X,1)).
\end{verbatim}
\newpage
\begin{verbatim}
State(S), PID(X), X != 0, $TRUE(barrier_out_arrived(S,X)) ->
State(assign_client_return(send_message
                           (receive_message(S,X),next(X),barrier_out),X,1)).

%
-> barrier_out_arrived([],I) = false.
-> barrier_out_arrived([PS(CLI_IN,CLI_OUT,MSG,Q)|T],I) =
   $IF(I == 0,
       first(Q) == barrier _out,
       barrier_out_arrived(T,I-1)).

-> assign_client_return([],I,V) = [].
-> assign_client_return([PS(CLI_IN,CLI_OUT,MSG,Q)|T],I,V) =
   $IF(I == 0,
       [PS(CLI_IN,V,MSG,Q)|T],
       [PS(CLI_IN,CLI_OUT,MSG,Q)|assign_client_return(T,I-1,V)]).

-> dequeue([]) = [].
-> dequeue([H|T]) = T.

-> receive_message([],I) = [].
-> receive_message([PS(CLI_IN,CLI_OUT,MSG,Q)|T],I) = 
   $IF(I == 0,
       [PS(CLI_IN,CLI_OUT,MSG,dequeue(Q))|T],
       [PS(CLI_IN,CLI_OUT,MSG,Q)|receive_message(T,I-1)]).
\end{verbatim}}

The model is 
based on the same simplification as the \promela\ model of the
algorithm: clients are not explicitly represented.  A valid system
state consists of three process states.  Each process state {\tt PS}
is a 4-tuple, whose every element implicitly corresponds to either one
of the variables used by the algorithm or the input queue.  In this
model the process state template is

\begin{verbatim}
      PS[client_barrier_in, client_barrier_out,
         holding_barrier_in, input_queue].
\end{verbatim}

In the initial state of each process, every variable has a zero value,
and the input queue is empty. The two transition rules correspond to
the handling operations for the \verb2barrier_out2 message, as also
shown in 
the \promela\ model above.   The rules are applicable when the client
barrier requests have been received by all processes, that is, the
\verb2client_barrier_in2 bit has been set by every process, any
\verb2holding_barrier_in2 bits that might have been set are again
reset.  In this formalization, the manager with 0 rank is the last to
release its client, so the state of the system to which the first of
the two state transition rules applies is

\begin{verbatim}
             (PS[1,0,0,(barrier_out)],
              PS[1,1,0,[]],
              PS[1,1,0,[]]),
\end{verbatim}

\noindent 
and the result of the application of the rule, including evaluation of
the functions 
\verb2assign_cli-2 \verb2ent_return2 to assign the value of the
\verb2client_barrier_out2 bit and \verb2receive_message2 to remove a
message from the input queue, is the overall system state 

\begin{verbatim}
                 (PS[1,1,0,[]],
                  PS[1,1,0,[]],
                  PS[1,1,0,[]]).
\end{verbatim}

Formalization of MPD algorithms in the sequent language of \otter\ is
similar to formalizations in the input language of the Mur$\phi$
model checker, especially as presented in the simplified form in
\cite{murphi:format}.  Similarities are most evident in the definition
of state transition rules.

\section{Results and Comparisons}
\label{sec:results}

At this point in the project the goal is to discover whether existing
tools and methodologies can be used for verification of MPD
algorithms.  We are also interested in developing formalization
techniques that simultaneously produce models abstract enough to allow
us to verify models of meaningful size and detailed enough to enable
automated model construction and/or code generation.  We apply
verification approaches to MPD algorithms that have been developed
some time ago, and are well understood and well debugged.  That 
is, bug hunting is currently only a secondary  goal. (We have
nonetheless found a minor error in the ring insertion algorithm using
the literal formalization of our earlier approach to using \spin.)
Rather, we are interested in developing an arsenal of verification
techniques to use for the analysis of the recent and future MPD 
algorithms.  Therefore, criteria for evaluation of verification
methods include the ease of modeling, correlation of the model to the
design and/or implementation of the MPD algorithm, and verification
performance.   

\subsection{Comparison of Formalizations}

The \spin/\promela\ approach is the more natural one for the MPD
application, but the \otter\ approach does offer some, albeit possibly
subjective, advantages. 

\spin\ and its input language \promela\ are specifically developed for
verification of concurrent communicating processes.  Therefore, the
language and the tool include special built-in constructs and
algorithms for message handling, communication path definition, and
variable declaration and manipulation.  As a result, there is a
natural mapping between MPD components and \promela\ entities.  By
contrast, \otter, as a general-purpose tool, has to be programmed from
scratch to handle these common operations of MPD algorithms.
Although, once constructed, this auxiliary portion of the \otter\
model can be reused, 
the \spin\ models are much more concise, as demonstrated
by the model extracts above.

From our point of view, the ability to  model an algorithm as a set of
state transition rules is the main advantage of the \otter\ approach.
The set of rules, which is the main component of the \otter\ model,
can be easily extracted from a flow chart representation of the MPD
algorithms or from other notations typically used in the early stages
of algorithm design.  Construction of \promela\ models looks and feels
much more like constructing another implementation of the algorithm,
only using a language much less powerful than C or Python, which are
used for actual implementations.  Furthermore, all intermediate states
produced by the \otter\ search are transparent to the user and can be
examined.  Such examination is not possible in \spin\ and may be a
limitation when trying to understand a particularly complicated error
trace.  

Correctness properties of \otter\ models of MPD algorithms have to be
formulated within the confines of first-order logic.  \spin, on the
other hand, allows one to record correctness properties in linear
temporal logic.  Although we have not yet encountered a situation where
limitations of first order logic prevented us from stating the desired
correctness property, such a situation is conceivable and may prove a
disqualifying drawback of the \otter\ approach.

\subsection{Performance Comparison}

All verification runs were conducted on a 933 MHz Pentium~III
processor with 970 MB of usable RAM.  We used default \spin\
settings for all verification attempts, except when we increased the
memory limit to allow the search to complete. In cases where
verification did not complete with default parameters within physical
memory limits, verification with compression (using \spin's
\texttt{-DCOLLAPSE} compile-time directive) was performed.  The input
files for the \otter\ experiments included settings to optimize the
state space exploration.  Some default flags that are usually needed
in standard \otter\ experiments but irrelevant for our application
were turned off.  

Table~\ref{tbl:stats} shows performance statistics of applying \spin\
\begin{table*}[t]
\caption{Verification statistics}
\vspace{.2in}
\centerline{
\begin{tabular}{lc|c|c|c|c|c} 
               &        & Model & Time & Memory & States & States \\
Problem        & Method & Size  & (s)  & (MB)   & Stored & Matched \\\hline
Unordered ring & Otter & 7  & 3979.08 & 391 & 7.83654e+05 & 1.908330e+06  \\ 
               & Spin  & 7  & 30.32  & 375 & 2.27008e+06 &1.90833e+06 \\ 
\hline
Ordered ring   & Otter & 14 & 15676 & 561 & 1.028351e+06 & 6.318042e+06 \\ 
               & Spin  & 14 & 263.68  & 734 & 6.50332e+06 & 8.66146e+06 \\ 
\hline
Barrier        & Otter & 19 & 27082 & 512 & 1.048593e+06 & 8.912898e+06\\
               & Spin  & 21 & 2045 & 746 & 8.38865e+06 & 8.38861e+07
\end{tabular}}
\vspace{0.1in}
    \label{tbl:stats}
\end{table*}
and \otter\ to three problems.  Two problems are variants of the ring
insertion algorithm.  In both versions the first daemon establishes a
singleton ring.  In the unordered version the subsequent daemons may
enter the ring in any order, resulting in many possible final
topologies with respect to the relative position of the processors in
the ring, and hence a much greater state search space and poorer
verification performance.  In the ordered version, the order in which
the processes enter the ring is fixed, resulting in a single possible
final topology.  In essence, the difference between the two versions
is that in the unordered case the daemons are numbered {\it before}
they begin to enter the ring, whereas in the ordered case they are
numbered {\it after} they have done so.  As a result, in the ordered
case, the processor that enters the ring first is always numbered with
one, the processor that is second, with two, and so on.  Because the
algorithm is independent of the identities of the processors, the
ordered version is less complex. 

Table 1 shows statistics for the largest model sizes on which a
particular verification approach succeeded.  On complex algorithms,
such as ring insertion, \otter\ matches \spin\ with respect to the
largest verifiable model.  In terms of speed and memory usage per
examined state, \otter\ is far behind \spin.  This result is not
surprising because \spin\ is a special-purpose tool, specifically
designed for applications like ours, while our use of \otter\ is
unusual in this case.  In fact, the performance of \otter\ far
exceeded our expectations.   

\otter\ could not verify models of as many processors as \spin\ could
for thr barrier algorithm.   The explanation for such a difference in
performance lies in the way states are represented in each method.
In \spin, the three local variables that contribute to individual
process states are bits, and the communications channels are
essentially arrays of bits, which allows \spin\ to store state vectors
very efficiently.  In addition, in the \spin\ verification run on a
model of twenty-one processors, state vectors were compressed,
resulting in further improved performance. In \otter, the
variables and the input queue are terms, which are not stored as
efficiently as bits.  Thus, \otter\ requires much more memory to store
an individual system state.  As a result, \otter\ is able to examine a
much smaller state spaces.

The \otter\ search engine is not optimized for the kind of search that
takes place in this application.  It is therefore not surprising that
verification with \otter\ is several times slower than with
\spin. But, since memory, not time, is the main limitation in this
application specifically and in model checking in general, the speed
of verification is only of minor concern.

\subsection{A Note on Model Sizes}
\label{sec:size-note}
Scalability was one of the design goals of MPD.  The daemon is
intended to run on hundreds, and eventually on thousands, of
processors.  Unexpected limitations of the underlying operating
systems aside, it is assumed that the same algorithms that execute
successfully in a daemon of just a few processors will execute
smoothly in a
much larger system.  Given the current state of the model-checking
technology, it is impossible to formally verify the algorithms of MPD
on very large models sizes, even if it were desirable to do so.
Luckily, the empirical evidence obtained while debugging the MPD code
using the traditional means suggests that the errors, even the most
difficult and obscure ones, exhibit themselves in daemons of just a few
processors.  It also appears to be an accepted view of the
model-checking community that to verify systems with unbounded
potential number of elements, it is sufficient to verify a limited( with
respect to the number of elements) model of the system
\cite{inca-compare,dill-protocol-hardware-92,jackson-counter-detector}.  

Our goal is to devise a verification approach and a modeling
methodology that allows us to verify complex MPD algorithms and
interactions of these algorithms on models of ten to twenty daemons.
For example, MPD contains an algorithm that merges two daemons that
are running parallel jobs.  We want to verify the algorithm on a
meaningful model consisting of two MPD structures, as shown
in Figure~\ref{fig:mpds-all}, each having a console process and
three of each of daemon, manager, and client processes.

\section{Summary}
\label{sec:summary}

We described here two approaches to verification of the algorithms of
the parallel process management system called MPD.  One approach is
based on the software model checker \spin, the other on the
general-purpose first-order theorem prover \otter.  Both approaches
are based on model checking, and the use of \otter\ in the model
checking-capacity is unusual.  The aim was to model algorithms of MPD
in both the \spin\ and the \otter\ approaches so as to enable 
verification of the largest possible model.  

The two approaches were compared with respect to the ease of
formalization and verification performance characteristics.  Overall,
\spin\ is more efficient in terms of absolute time and memory
requirements and relative time and memory requirements per stored
system state.  In terms of the size of models that each approach
allows us to verify, both tools perform roughly the same, with \spin\
occasionally demonstrating better performance.  Neither approach allows
us to verify complicated algorithms on models of about twenty daemons,
which is our goal.  In terms of formalization methodology, the two
approaches are too different to compare, and both exhibit
advantages and disadvantages.  

The main goal of this technical note is to document the current
approach to modeling MPD in \promela, and to describe how \otter\ can
be used to simulate model-checking style search.  We have been
applying model checking techniques to MPD algorithms that have been
under development and testing for some time.  Consequently, it should
come as no surprise that no errors have been discovered.  Even though
the current modeling methodologies limit verification to models of
only a few entities, applying them to new MPD algorithms could still
be beneficial.

For further details the reader is referred to the
complete models of the presented algorithms.  This information is
available at \texttt{http://www.mcs.anl.gov/\~{ }matlin/spin-mpd}.

\addcontentsline{toc}{section}{References}
\bibliographystyle{plain}
\bibliography{master}

\begin{thebibliography}{10}

\bibitem{inca-compare}
G.~S. Anrunin, J.~C. Corbett, M.~B. Dwyer, C.~S. Pasareanu, and S.~F. Siegel.
\newblock Comparing finite-state verification techniques for concurrent
  software.
\newblock Technical Report UM-CS-1999-069, Department of Computer Science,
  University of Massachusetts, 1999.
\newblock http://ext.math.umass.edu/\~{ }avrunin/recent\_pubs/comparing.ps,
  Under revision 2002.

\bibitem{bgl00:mpd:pvmmpi00}
R.~Butler, W.~Gropp, and E.~Lusk.
\newblock A scalable process-management environment for parallel programs.
\newblock In J.~Dongarra, P.~Kacsuk, and N.~Podhorszki, editors, {\em Recent
  Advances in Parallel Virutal Machine and Message Passing Interface}, LNCS
  1908, pages 168--175. Springer Verlag, September 2000.

\bibitem{butler-lusk-gropp:mpd-parcomp}
R.~Butler, W.~Gropp, and E.~Lusk.
\newblock Components and interfaces of a process management system for parallel
  programs.
\newblock {\em Parallel Computing}, 27:1417--1429, 2001.

\bibitem{dill-protocol-hardware-92}
D.~L. Dill, A.~J. Drexler, A.~J. Hu, and C.~H. Yang.
\newblock Protocol verification as a hardware design aid.
\newblock In {\em 1992 IEEE International Conference on Computer Design: VLSI
  in Computers and Processors}, pages 522--525. IEEE Computer Society, 1992.
\newblock Cambridge, MA, October 11-14.

\bibitem{gropp-lusk:mpich-www}
W.~Gropp and E.~Lusk.
\newblock {MPICH}.
\newblock ftp://info.mcs.anl.gov/pub/mpi.

\bibitem{gropp-lusk-doss-skjellum:mpich}
W.~Gropp, E.~Lusk, N.~Doss, and A.~Skjellum.
\newblock A high-performance, portable implementation of the {MPI}
  {M}essage-{P}assing {I}nterface standard.
\newblock {\em Parallel Computing}, 22(6):789--828, 1996.

\bibitem{spin:book}
G.~J. Holzmann.
\newblock {\em Design and Validation of Computer Protocols}.
\newblock Prentice Hall, 1991.

\bibitem{spin:article}
G.~J. Holzmann.
\newblock The model checker {SPIN}.
\newblock {\em IEEE Transactions on Software Engineering}, 22(5):279--295, May
  1997.

\bibitem{murphi:format}
C.~N. Ip and D.~L. Dill.
\newblock Better verification through symmetry.
\newblock {\em Formal Methods in System Design}, 9(1--2):41--75, 1996.

\bibitem{jackson-counter-detector}
D.~Jackson and C.~A. Damon.
\newblock Elements of style: Analysing a software design feature with a
  counterexample detector.
\newblock {\em I{E}{E}{E} Transactions on Software Engineering},
  22(7):484--495, July 1996.

\bibitem{acl2:book}
M.~Kaufmann, P.~Manolios, and J~S. Moore.
\newblock {\em Computer-Aided Reasoning: An Approach}.
\newblock Kluwer Academic Publishers, 2000.

\bibitem{spin-mpd}
O.~S. Matlin, E.~Lusk, and W.~McCune.
\newblock {SPIN}ning parallel systems software.
\newblock In D.~Bosnacki and S.~Leue, editors, {\em Model Checking Software.
  Proceedings of the 9th International SPIN Workshop}, LNCS 2318, pages
  213--220. Springer Verlag, 2002.

\bibitem{McCune94a}
W.~McCune.
\newblock Otter 3.0 {R}eference {M}anual and {G}uide.
\newblock Tech. Report ANL-94/6, Argonne National Laboratory, Argonne, IL,
  1994.
\newblock See also URL http://www.mcs.anl.gov/AR/otter/.

\bibitem{mpd-acl2}
W.~McCune and E.~Lusk.
\newblock {ACL2} for parallel systems software.
\newblock In M.~Kaufmann and J~S. Moore, editors, {\em Proceedings of the 2nd
  {ACL2} Workshop}. University of Texas, 2000.
\newblock http://www.cs.utexas.edu/users/moore/acl2/workshop-2000.

\bibitem{mpi-forum:journal}
{Message Passing Interface Forum}.
\newblock {MPI}: A {M}essage-{P}assing {I}nterface standard.
\newblock {\em International Journal of Supercomputer Applications},
  8(3/4):165--414, 1994.

\bibitem{mpi-forum:mpi2-journal}
{Message Passing Interface Forum}.
\newblock {MPI2}: A {M}essage {P}assing {I}nterface standard.
\newblock {\em International Journal of High Performance Computing
  Applications}, 12(1--2):1--299, 1998.

\bibitem{acl2:oracle}
J~S. Moore.
\newblock A mechanically checked proof of a multiprocessor result via a
  uniprocessor view.
\newblock http://www.cs.utexas.edu/users/moore/publications/acl2-papers.html,
  February 1998.

\bibitem{spin:low-fat}
T.~C. Ruys.
\newblock Low-fat recipes for {SPIN}.
\newblock In K.~Havelund, J.~Penix, and W.~Visser, editors, {\em Proceedings of
  the 7th International SPIN Workshop}, LNCS 1885, pages 287--321. Springer
  Verlag, 2000.

\bibitem{phd:ruys}
T.~C. Ruys.
\newblock {\em Towards Effective Model Checking}.
\newblock PhD thesis, University of Twente, 2001.

\bibitem{spin-web-page}
Spin home page.
\newblock http://spinroot.com.

\bibitem{stevens-unp1}
W.~R. Stevens.
\newblock {\em Unix Network Programming}, volume~1.
\newblock Prentice Hall {PTR}, second edition, 1998.

\end{thebibliography}

\end{document}